# Thermal Properties of the Binary-Filler Composites with Few-Layer Graphene and Copper Nanoparticles

Zahra Barani[1,2], Amirmahdi Mohammadzadeh[2,3], Adane Geremew[1,2], Chun Yu Tammy Huang[2,3], Devin Coleman[3], Lorenzo Mangolini[3,4], Fariborz Kargar[1,2,*], and Alexander A. Balandin[1,2,3,§]

[1]Department of Electrical and Computer Engineering, Bourns College of Engineering, University of California, Riverside, California 92521 USA

[2]Phonon Optimized Engineered Materials (POEM) Center, University of California, Riverside, California 92521 USA

[3]Materials Science and Engineering Program, Bourns College of Engineering, University of California, Riverside 92521 USA

[4]Department of Mechanical Engineering, Bourns College of Engineering, University of California, Riverside 92521 USA

[§] Corresponding authors: fkarg001@ucr.edu and balandin@ece.ucr.edu ; web-site: http://balandingroup.ucr.edu/

## Abstract

The thermal properties of an epoxy-based binary composites comprised of graphene and copper nanoparticles are reported. It is found that the "synergistic" filler effect, revealed as a strong enhancement of the thermal conductivity of composites with the size-dissimilar fillers, has a well-defined filler loading threshold. The thermal conductivity of composites with a moderate graphene concentration of $f_g = 15$ wt% exhibits an abrupt increase as the loading of copper nanoparticles approaches $f_{Cu}{\sim}40$ wt%, followed by saturation. The effect is attributed to intercalation of spherical copper nanoparticles between the large graphene flakes, resulting in formation of the highly thermally conductive percolation network. In contrast, in composites with a high graphene concentration, $f_g = 40$ wt%, the thermal conductivity increases linearly with addition of copper nanoparticles. The electrical percolation is observed at low graphene loading, $f_g < 7$ wt.%, owing to the large aspect ratio of graphene. At all concentrations of the fillers, below and above the electrical percolation threshold, the thermal transport is dominated by phonons. The obtained results shed light on the interaction between graphene fillers and copper nanoparticles in the composites and demonstrate potential of such hybrid epoxy composites for practical applications in thermal interface materials and adhesives.

## 1. Introduction

The growth in power densities in electronic, optoelectronic and microwave devices makes efficient thermal management a critical issue[1–3]. The development of the next generation of thermal interface materials (TIMs) with substantially higher thermal conductivity is essential for various device technologies. The state-of-the-art light emitting diodes[4], lithium-ion batteries[5,6], and solar cells[7] suffer from the inadequate heat conduction properties of commercial TIMs, which include thermal greases, adhesives and thermal phase-change materials. The polymer-based TIMs, applied between heat sources and heat sinks, are vital components of passive heat management in electronic systems. Polymers, in general, due to their amorphous structure, possess low thermal conductivity in the range of $0.2-0.5\ \mathrm{Wm^{-1}K^{-1}}$[8]. The conventional approach for producing TIMs is incorporation of microscale and nanoscale particles into the polymer matrix in order to enhance the overall heat conduction properties of the resulting composites[9]. The filler particle can be metallic or electrically insulating. In some applications, one desires to increase thermal conductivity while preserving electrical insulation, in other applications, one benefits from increasing both thermal and electrical conductivity. The commercially available TIMs, with complex preparation recipes, have the "bulk" thermal conductivities in the range of $0.5-5\ \mathrm{Wm^{-1}K^{-1}}$, achieved at high filler loading fractions of $\phi\sim 50$ vol%[10]. To satisfy the industry needs one need to develop novel TIMs, including cured epoxies and non-cured thermal greases, with the "bulk" thermal conductivity in the range of 15–25 $\mathrm{Wm^{-1}K^{-1}}$ near room temperature (RT). Substantial increase in the heat conduction properties of composites requires detailed investigation of the alternative heat conduction fillers and better understanding of the filler – filler and filler – matrix interactions.

Graphene has attracted a lot of attention owing to its extraordinary electrical[11,12], optical[13], and thermal properties[14–20]. The *intrinsic* thermal conductivity of a large sheet of single-layer graphene (SLG) can exceed than that of the basal planes of high quality graphite, which by itself is high: ~$2000\ \mathrm{Wm^{-1}K^{-1}}$ at RT[10,14,21–23]. Although the few-layer graphene (FLG) sheets have lower intrinsic thermal conductivity than SLG, they are more technologically feasible for practical applications[10,24]. The FLG fillers preserve the high thermal conductivity, close to that of graphite, and possess mechanical flexibility, facilitating coupling to the matrix material. It is known that FLG can be mass produced at low cost, which is an important factor for any fillers in TIMs, composites, and coatings [2,5,7,10,24–29]. The sheets of FLG are also less vulnerable to defects induced by processing, mechanical stresses, rolling, and folding, which happens often during the mixing of fillers with the polymer matrix[10,24,25]. The first study of graphene composites reported an enhancement of the thermal conductivity of epoxy from 0.2 $\mathrm{Wm^{-1}K^{-1}}$ to $\sim 5\ \mathrm{Wm^{-1}K^{-1}}$ at the low graphene loading of $\sim 10$ vol% [10]. The results have been independently confirmed and improved by other research groups, which obtained similar enhancement factors at lower graphene concentrations [30,31].

The first graphene-based composites have been prepared using raw graphite as the source material to produce SLG and FLG fillers via liquid phase exfoliation (LPE). The preparation method includes chemical processing, sonication and centrifugation [10]. These procedures do not allow for an accurate control of the lateral dimensions and thicknesses of the resulting graphene fillers. However, a certain range in the size distribution of FLG fillers turned out to be even beneficial. The size distribution of the fillers can result in the “synergistic” effect, characteristic for the fillers of different dimensions [31–39]. The “synergistic” effects constitutes a stronger enhancement of the thermal conductivity of the composites with the size-dissimilar binary fillers than with the individual fillers of the same total concentration [31–39]. Recent technological advancements made it possible to produce FLG fillers of different sizes and thicknesses using liquid-phase exfoliation (LPE)[40,41] or graphene oxide reduction methods [42–45]. These developments open up a possibility of industrial production of composites with the high loading of graphene[24]. Here and below, in the thermal context, we use the term graphene to indicate a mixture of SLG and FLG.

The physical nature of heat transport is different in metallic and non-metallic fillers. In non-magnetic solid materials, heat is carried by electrons and phonons – quanta of the ion-core crystal lattice vibrations. The thermal conductivity of solid materials is described as $\lambda = \lambda_p + \lambda_e$ where $\lambda_p$ and $\lambda_e$ are the phonon and electron contributions, respectively. In electrical insulators and semiconductors, heat is mostly transferred by acoustic phonons with long mean free paths (MFP)[16]. The same is true for graphene and other carbon allotropes[16]. An equation for $\lambda_p$, based on the gas kinetics, is $\lambda_p = (1/3)c_p \upsilon \Lambda$, where $c_p$ is the specific heat, $\upsilon$ is the phonon's average group velocity, which in many solids can be approximated by the sound velocity, and $\Lambda$ is the phonon MFP, respectively. The average *grey* MFP of acoustic phonons in graphene is ~750 nm at RT[16,22]. For this reason, in thermal applications, it is favorable to use graphene fillers with comparable or large lateral dimensions, exceeding the phonon MFP in order to avoid degradation of the intrinsic thermal conductivity of the fillers due to the phonon – filler edge scattering. It should be noted that using excessively large graphene fillers can lead to the agglomeration of the fillers during mixing as well as filler bending and rolling. There exists optimum lateral size and thickness ranges for FLG fillers for each specific matrix material and desired characteristics of the composites. In metals, the heat conduction is dominated by electrons due to their high concentration. For example, pure copper has thermal conductivity of $\sim 400\ \mathrm{Wm^{-1}K^{-1}}$ in which electrons contribute ~98% of the total thermal conductivity at RT [16]. Electrons have much shorter average MFP compared to acoustic phonons. In pure copper, the electron MFP is ~40 nm[46]. For this reason, one can envision using metallic fillers of much smaller dimensions without degrading their intrinsic heat conduction properties.

Here, we report the results of the investigation of the thermal conductivity of the epoxy composites with hybrid fillers comprised of FLG with the large lateral size (few μm) and copper nanoparticles (Cu-NP) with the small lateral size (few nm). In the high-loading composites with FLG and Cu-NP fractions of $f_g = 40$ wt% and $f_{Cu} = 35$ wt%, we achieved $13.5 \pm 1.6\ \mathrm{Wm^{-1}K^{-1}}$ thermal conductivity, which translates into ~6750% enhancement of the polymer's thermal conductivity. It has been established that the increase in the Cu-NP loading, at constant graphene concentration, results in incorporation of copper nanoparticles between the large graphene flakes, with the corresponding formation of the thermally conductive network of fillers. The thermal transport in the thermal percolation regime is characterized by a significant enhancement in thermal conductivity. Our results show that a combination of graphene fillers (high-aspect ratio and μm-scale lateral dimensions) with the phonon-dominated heat conduction and metallic fillers (small aspect ratio and nm-scale dimensions) with the electronic heat conduction is promising for TIM applications.

## 2. Results and Discussion

In Figure 1a-g) we illustrate the step-by-step preparation procedures for the samples. The base polymer material is a curing epoxy composed of a resin (Bisphenol-A; Allied High Tech Products, Inc.) and a hardening agent (Triethylenetetramine; Allied High Tech Products, Inc.). Commercially available copper nanoparticles (US Research Nanomaterials, Inc.) with the average diameter of 100 nm are mixed with the resin in pre-calculated proportions inside a glove box in the argon gas atmosphere (Figure 1 (a)). The level of oxygen inside the glove box is carefully monitored and kept below ~0.2 ppm to prevent oxidation of Cu-NPs. It should be noted that Cu-NPs like other metallic nanoparticles are highly flammable and tend to oxidize upon exposure to air. The oxidation of nanoparticles reduces the thermal conductivity of copper at least one order of magnitude[47]. In the next step, graphene fillers (graphene, grade H, XG-Sciences) with the vendor specified original lateral dimensions of ~25 μm are weighed and added to the Cu-NP-resin mixture outside the glove box. Figure 1 (b) shows a representative scanning electron microscopy (SEM) image of graphene flakes confirming their large lateral dimensions. At the next step, the hardening agent is added and mixed with the rest of the fillers and resin using a high shear speed mixer (Flacktek Inc.). The mixture is vacuumed for ~10 minutes to extract possible gas bubbles, which are trapped inside the solution during the mixing process (Figure 1 (c)). This process is repeated several times in order to achieve a uniform dispersion and minimize the voids inside the composite. The samples are poured in silicon molds, lightly pressed, and left inside the oven for ~2 hours at 70 ºC to cure and solidify (Figure 1 (d)). All composites are prepared in the form of disks with a diameter of 25.6 mm and thickness of ~1 mm. An optical image of a highly loaded composite with $f_g = 15$ wt% and $f_{Cu} = 40$ wt% is shown in Figure 1 (e). In Figure 1 (f) we show an SEM image of the cross section of the same sample, which clearly demonstrates that the smaller Cu nanoparticles tend to reside between the large graphene fillers. The latter helps in creation of the thermal percolation networks, illustrated in Figure 1 (g). In this image, the large hexagons and small spheres represent FLG and Cu-NPs, respectively. The red arrows show the highly conductive heat transport paths through fillers inside the epoxy polymer host.

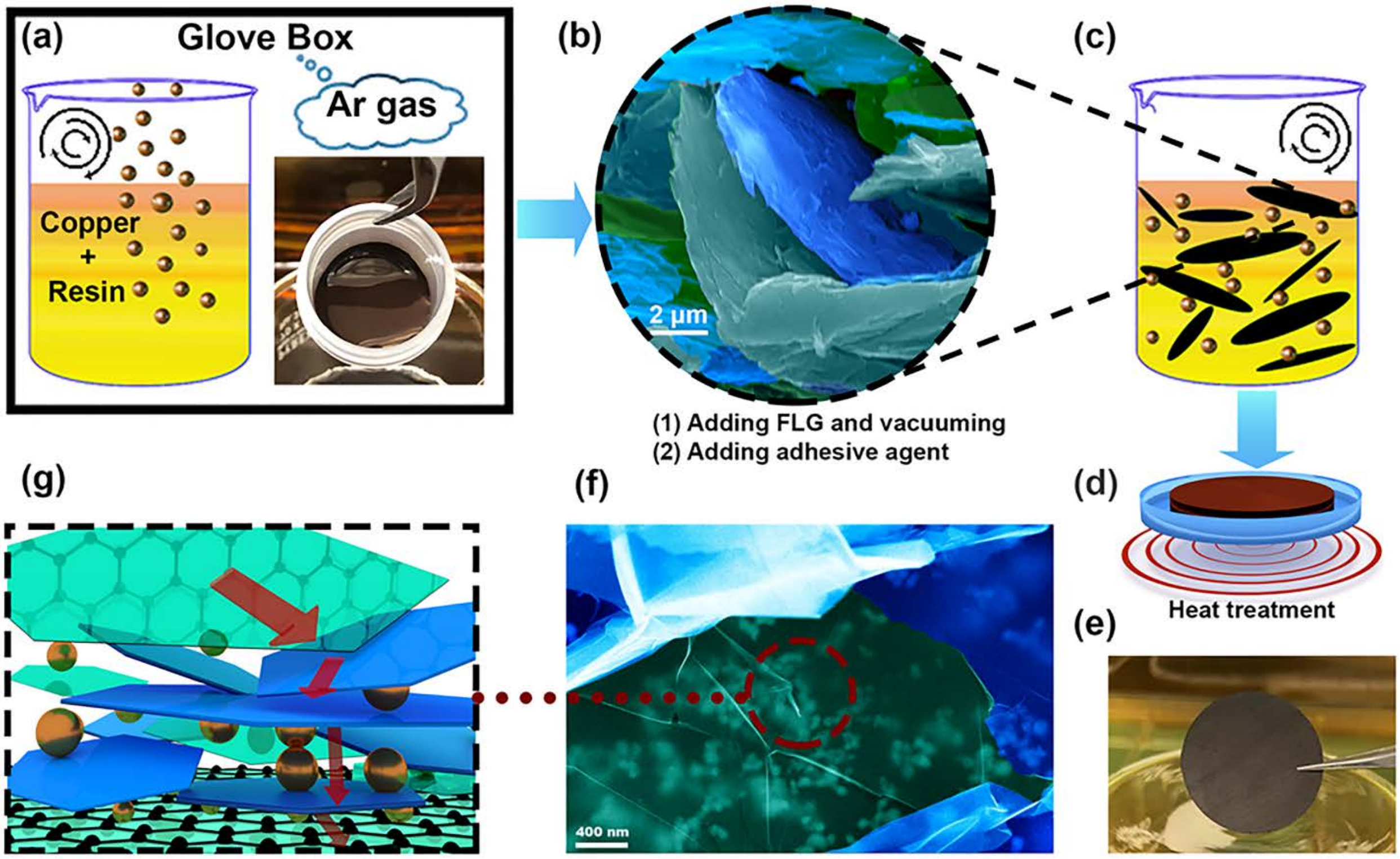

**Figure 1:** Schematic of the composite preparation and sample characterization. (a) Addition of Cu-NPs to the epoxy resin and mixing them inside the glove box in argon gas environment in order to prevent oxidation of Cu-NPs. (b) Scanning electron microscopy (SEM) image of large FLG fillers with the lateral dimensions in the range of ~15 – 25 µm. (c) Mixing FLG fillers and hardener with the prepared resin – Cu-NP mixture and vacuuming. (d) Curing the mixture in the temperature controlled oven for ~2 hours. (e) Optical image of a highly loaded composite with 15 wt% of FLG and 40 wt% Cu-NP fillers. (f) SEM image of the cross section of the same sample demonstrating the overlapping of FLG fillers and intercalation of Cu nanoparticles between them. (g) Schematic of the thermal percolation network of the fillers. The Cu-NPs bridge between the FLG fillers and create thermal paths, which significantly enhance the thermal transport via the highly conductive fillers.

An accurate measurement of the mass density of the composites is important in order to determine thermal conductivity from the measurement of the thermal diffusivity. The mass density of the samples was measured using an electronic scale utilizing Archimedes' principle. The mass density was then utilized to calculate the porosity of the samples and determine the thermal conductivity, $\lambda$, by the transient "laser flash" method (LFA) method[48,49]. We measured the specific heat and thermal diffusivity of the samples with the LFA instrument and calculated the cross-plane thermal conductivity according to $\lambda = \rho_c c_p \alpha$ where $\rho_c$, $c_p$, and $\alpha$ are composite's density, specific heat, and cross-plane diffusivity, respectively. The details of the density and LFA thermal diffusivity, conductivity, and specific heat measurements are provided in the Experimental Section. Figures 2

(a) and (b) show the mass density and specific heat of the prepared samples, respectively. The data are presented for the binary fillers at constant graphene loadings of 5 wt%, 15 wt%, and 40 wt% and various copper nanoparticle loading fractions.

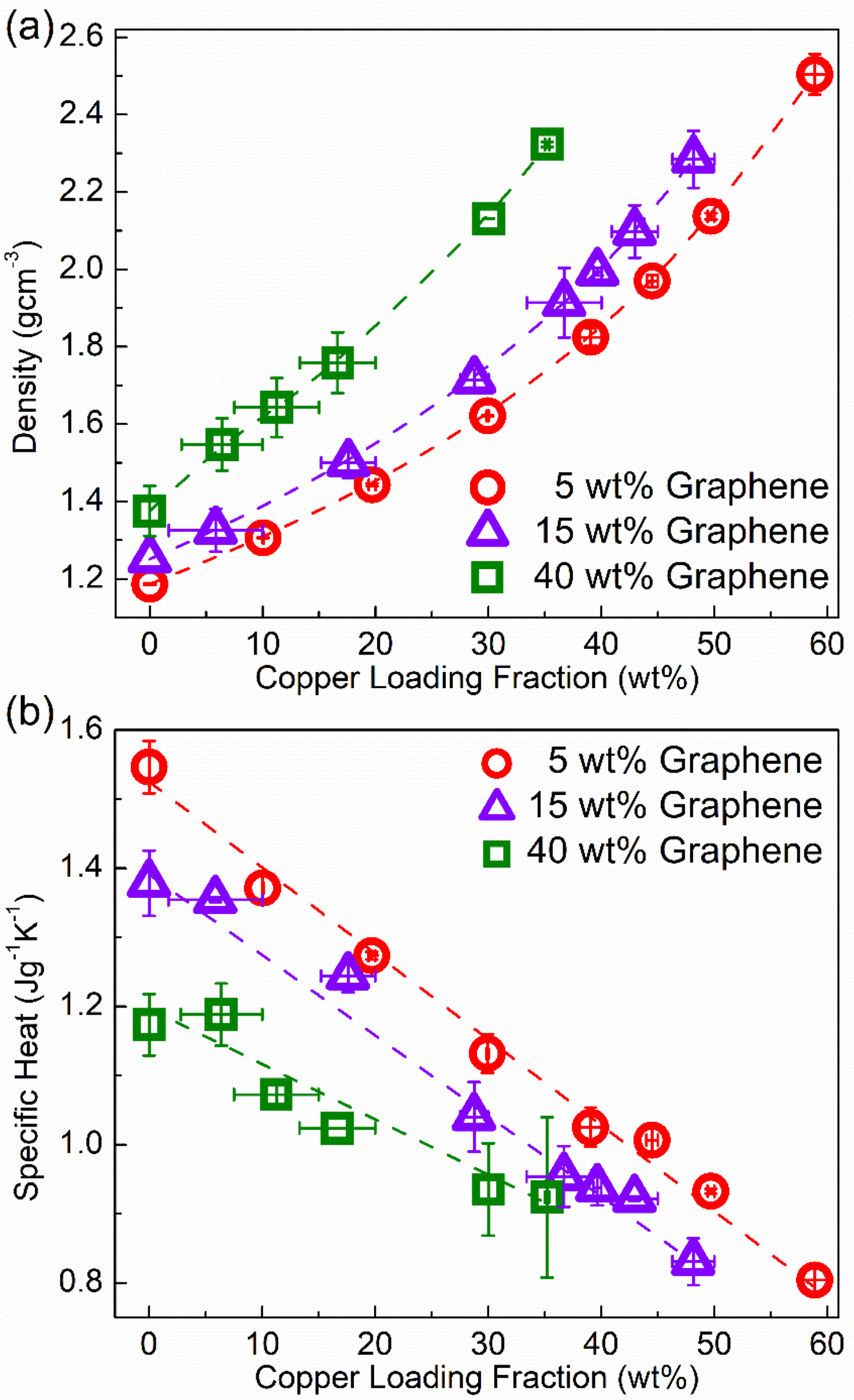

**Figure 2:** (a) Mass density, and (b) specific heat of the hybrid composites at constant mass loading fraction of graphene as a function of Cu-NP loading. The dashed lines show the fitting of the thermos-physical properties of the composites based on the rule of mixtures, which indicates an excellent agreement with the experimental data. While the density of the composites increases non-linearly with addition of copper nanoparticles, the specific heat decreases linearly as a function of the copper mass loading fraction.

The results of the mass density measurements presented in Figure 2 (a) indicate that addition of Cu-NP fillers to the composites with the constant graphene content, the density of the samples increases. This is because Cu has a much larger density compared to graphene ($\rho_{Cu}/\rho_g$~4) and pure epoxy ($\rho_{Cu}/\rho_e$~7.8). More importantly, the density data confirms that the porosity of the samples is less than 7% of the composite total volume, even at high loading fractions. The porosity is calculated as $\beta = (\rho_{th} - \rho_{exp})/\rho_{th}$, where $\rho_{exp}$ is the measured density of the actual sample and $\rho_{th}$ is the theoretical density of the composite according to the rule of mixture, defined as $\rho_{th} = \sum m_i / \sum (m_i/\rho_i)$ , where $m_i$ and $\rho_i$ are the mass and density of the composite constituents (epoxy, graphene, and copper nanoparticles). The excellent agreement between the experimental data and theoretical dashed lines confirms that our composites have a negligible fraction of air gaps. In contrast to the density, the specific heat of the composites decreases linearly with the addition of Cu-NP due to the fact that the specific heat of Cu is smaller than that of both graphene ($c_{p,Cu}/c_{p,g}$ ~ 0.53 ) and pure epoxy ($c_{p,Cu}/c_{p,e}$ ~ 0.25). In Figure 2 (a-b), the horizontal error bars include the uncertainties associated with the Cu-NP mass fraction in the samples determined from averaging over several samples. The vertical error bars are defined by the uncertainties in the experimental measurements, including the instrumental and standard deviation in multiple measurements.

The cross-plane thermal diffusivity of the composites was measured at RT using the LFA technique. Figure 3 (a) shows the thermal diffusivity of the composites at a constant graphene concentration as a function of Cu-NP loading. In order to analyze the data, we introduce three different thermal regimes determined by the concentration of the graphene filler: low ($f_g =$ 5 wt%), medium ($f_g = 15$ wt%), and high ($f_g = 40$ wt%) as illustrated with the red circles, violet triangles, and green squares, respectively. At low and medium graphene loading, the diffusivity of the composites increases linearly with addition of Cu-NP fillers up to a certain loading threshold. At the low graphene concentration regime, starting at $f_{Cu} = 40$ wt% Cu-NP loading, the diffusivity increases linearly at a higher rate. At the medium graphene loading regime, the diffusivity experiences an abrupt jump when the Cu-NP loading reaches $f_{Cu} = 40$ wt%. At this point the diffusivity increases by almost factor of ×2 and then saturates, *i.e.* adding more Cu-NP fillers does not affect the thermal diffusivity. For the high graphene loading fractions, adding Cu-NP fillers increases the diffusivity linearly without jumps or changes in the slope.

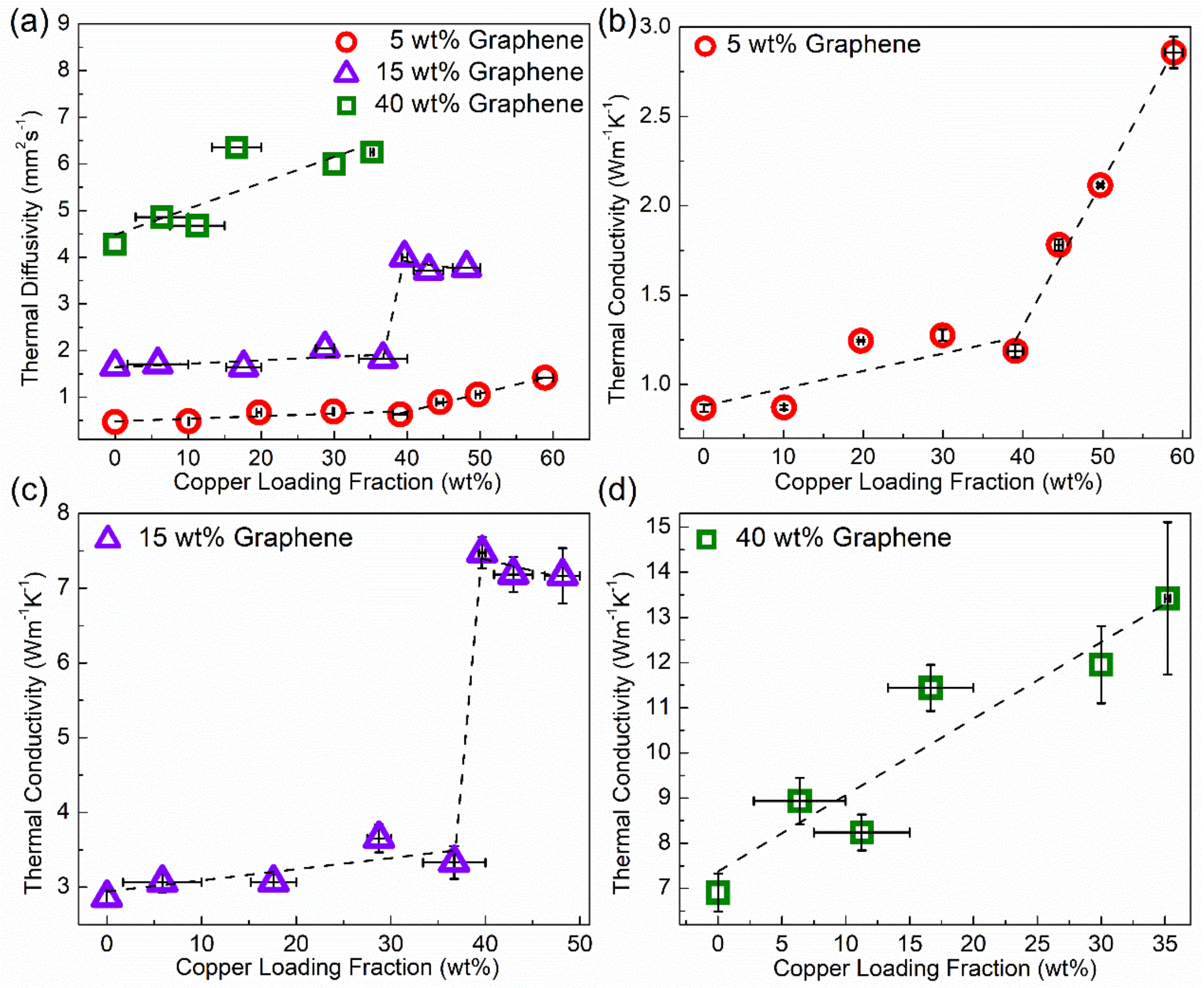

**Figure 3:** (a) Thermal diffusivity of the epoxy composites with binary fillers of graphene and copper nanoparticles as a function of copper mass loading fraction. The diffusivity of the composites with 5 wt% of graphene increases gradually with the copper loading. The diffusivity grows faster as the copper mass fraction exceeds 40 wt% due to creation of the thermally conductive filler paths inside the polymer host. In composites with 15 wt% of graphene and at ~35 wt% of copper, the thermal percolation reveals as an abrupt change in the thermal diffusivity. At 40 wt% of graphene loading, the system is already in the thermal percolation regime with the graphene fillers and thus, the diffusivity increases linearly with the addition of copper nanoparticles. Thermal conductivity of the epoxy with (b) 5 wt%, (c) 15 wt%, and (d) 40 wt% of graphene as a function of copper loading fraction.

Figures 3 (b-d) show the thermal conductivity of the composites determined from their measured thermal diffusivity (Figure 3 (a)), mass density (Figure 2 (a)) and specific heat (Figure 2 (b)). The thermal conductivity of the samples at all three graphene loadings (low, medium and high) follows the same trend as described for the thermal diffusivity. At the low graphene loading fraction, $f_g = 5$ wt%, graphene fillers are dispersed randomly in the polymer matrix and are separated apart, not forming a percolated network. Addition of the Cu-NP fillers increases the thermal conductivity slightly as expected from the effective medium considerations[24,50,51]. At a large loading of Cu-NP, $f_{cu}$~40 wt%, the copper nanoparticles fill in the gaps between the highly thermally conductive graphene fillers, and create a limited number of highly conductive thermal paths (see Figure 1 (g)), which results in steeper increase in the thermal conductivity (Figure 3 (b)). It should be noted that copper by itself has a rather high thermal conductivity of $\sim 400\ \mathrm{Wm^{-1}K^{-1}}$[52] although it is smaller than that of FLG fillers, which is on the order of ~$2000\ \mathrm{Wm^{-1}K^{-1}}$ at RT[14,53]. The change in the dependence of the thermal conductivity of composites with a single type of filler at high loading fractions has been discussed in literature previously[54]. Available models predict that the thermal conductivity of the polymer composites increases linearly with the increasing loading of the highly thermally conductive fillers. At certain filler loading fraction, defined as the thermal percolation threshold, a percolated network of highly conductive fillers forms inside the poorly conducting polymer matrix and enhances the thermal transport. The latter reflects in the change in the slope of the linear dependence. Our results for the low graphene loading fraction are in agreement with this previously described scenario.

We now turn to the composites with the medium graphene loading fraction, $f_g = 15$ wt%. Even before adding Cu-NP fillers, the thermal conductivity of such composite is relatively high, $\sim 3\ \mathrm{Wm^{-1}K^{-1}}$, which surpasses the thermal conductivity of many commercial TIMs with higher filler loadings. Addition of Cu-NP fillers to the samples, before reaching to the thermal percolation threshold, results in slow increase in the thermal conductivity, at a rate similar to the one in low graphene loading samples. At the Cu-NP loading of $\sim f_{cu}\sim 40$ wt%, the thermal conductivity reveals an abrupt increase due to reaching the thermal percolation threshold (Figure 2 (c)). Even though the thermal percolation is reached at about the same Cu-NP loading the thermal conductivity trend as a function of Cu-NP loading is different (compare Figure 2 (b) and (c)). The change from the below-percolation to percolation thermal transport regime is abrupt and revels the saturation behavior for high Cu-NP loading. The highest thermal conductivity limit at a given graphene concentration is reached. In other words, at $f_g = 15$ wt% and $f_{cu} = 40$ wt%, all possible percolation paths of highly conductive fillers including graphene-graphene, graphene-Cu-NP, and Cu-NP – Cu-NP have been formed. These results confirm the existence of an optimum loading fraction for each filler in composites with the binary dissimilar fillers, which has been reported in some studies for other types of fillers [31,32].

In the high graphene loading composites, $f_g = 40$ wt%, the thermal conductivity increases approximately linearly with addition of the Cu-NP fillers, and reaches $\lambda \sim 13.5 \pm 1.6\ \mathrm{Wm^{-1}K^{-1}}$ at $f_{Cu} = 35$ wt%. The rate of the thermal conductivity increase with Cu-NP loading is larger compared to the samples with the lower graphene loadings. This can be explained in the following way. The composites with the 40 wt% loading of large graphene fillers are already in the thermal percolation transport regime or close to it. Adding the Cu-NP fillers helps to connects the graphene fillers more effectively and thus, enhances the overall heat transport. Note that larger thermal conductivity data scatter is a signature of the thermal percolation regime as reported in the studies with other types of fillers[55]. Table I summarizes reported thermal properties of composites with the hybrid fillers, which revealed "synergistic" effects. For better comparison, the data are primarily shown for polymeric composites with various quasi-2D fillers, *e.g.* graphene and *h*-BN.

**Table I:** Thermal Conductivity of Composites with Hybrid Fillers

| Filler Type | TC | Loading | Matrix | Refs. |
|---|---|---|---|---|
| GNP/Cu-NP | 13.5±1.6 | 40 / 35 wt% | epoxy | This work |
| GNS / MWCNT (both silver functionalized) | 12.3 | total:20 vol.% | PVA | [56] |
| GNP / *h*-BN | 4.7 | 16 / 1 vol.% | epoxy | [31] |
| GNP / $Al_2O_3$ / MgO | 3.1 | 0.5 / 48.7 / 20.8 wt% | PC/ABS | [57] |
| CNTs grown on the GNP | 2.4 | 20 wt% | epoxy | [58] |
| GNP / *h*-BN | 1.8 | 20 / 1.5 wt% | PA | [59] |
| GNP / MWCNT | 1.4 | 18 / 2 wt% | PC | [60] |
| Ag NWs/ GNP (functionalized) | 1.4 | 4 vol% / 2 wt% | epoxy | [61] |
| GNP decorated with $Al_2O_3$ | 1.5 | 12 wt% | epoxy | [62] |
| Ag nanoparticle decorated GNS | 1.0 | 5 wt% | epoxy | [63] |
| GNP / *h*-BN (nanosheet) | 0.9 | 6.8 / 1.6 wt% | PA6 | [64] |
| GNP / *h*-BN | 0.7 | 20 / 1.5 wt% | PS | [59] |
| GNP / Ni | 0.7 | 5.0 /8 wt% | PVDF | [65] |
| GNP / MgO | 0.5 | 30 wt% | epoxy | [66] |
| MgO / GNP (coated) | 0.4 | 7 wt % | epoxy | [67] |
| GNP / MWCNT | 0.3 | 0.1 / 0.9 wt% | epoxy | [68] |
| GO / MWCNT | 4.4 | 4.64 / 0.36 wt% | epoxy | [69] |
| GO / AlN | 2.8 | 6 / 50 wt% | epoxy | [70] |
| $Al_2O_3$ / rGO (functionalized) | 0.3 | 30 / 0.3 wt% | epoxy | [71] |
| *h*-BN (vertically aligned) / SiC | 5.8 | 40 wt% | epoxy | [72] |
| 3D BN / rGO | 5.1 | 13.2 wt% | epoxy | [73] |
| *h*-BN / AlN (anisotropic/spherical) | 4.1 (in-plane) | 30 wt% | PI | [74] |
| *h*-BN (whiskers/aggregated particles) | 3.6 | 12.9 / 30.1 vol% | epoxy | [75] |
| *h*-BN (µm and nm size) | 2.6 | 40 /20 wt% | PPS | [33] |
| *h*-BN / MWCNT (Functionalized) | 1.9 | 30 / 1 vol% | epoxy | [76] |
| *h*-BN / MWCNT | 1.7 | 50 / 1 wt% | PPS | [77] |
| *h*-BN (µm/nm sized) | 1.2 | 30 wt% | PI | [34] |
| Ag nanoparticle-deposited BN | 3.1 | 25.1 | epoxy | [78] |
| MWCNT / micro – SiC (functionalized) | 6.8 | 5 / 55 wt% | epoxy | [79] |
| MWCNT / AlN | 1.0 | 4 / 25 wt% | epoxy | [80] |
| MWCNT / Cu | 0.6 | 15 / 40 wt% | epoxy | [81] |
| AlN / $Al_2O_3$ (large/small size) | 3.4 | 40.6 /17.4 wt% | epoxy | [32] |
| SNPs / Ag NWs | 1.1 | 40 / 4 wt% | epoxy | [82] |
| GO-encapsulated *h*-BN (*h*-BN@GO) | 2.2 | total: 40 wt% | epoxy | [83] |
| AlN (whiskers/spheres) | 4.3 | 30 / 30 vol.% | epoxy | [38] |
| GNS / CINAP | 4.1 | 5 / 15 wt% | CE | [84] |
| Cu and tin-zinc alloy microfibers | 2.3 | 25 / 19 vol.% | PA6 | [85] |
| CuNPs-CuNWs@BN | 4.3 | 10 wt% | PI | [86] |

The cross-plane electrical resistivity, $\zeta$, of the composites was measured using a standard two-probe configuration. The electrical conductivity, $\sigma = 1/\zeta$, was calculated for each sample and plotted as a function of graphene and Cu-NP loading fractions. The details of the electrical resistivity measurements are provided in the Experimental Section. Figure 4 (a) shows the cross-plane electrical conductivity of epoxy with graphene as a function of FLG loading fraction. The epoxies are electrically insulating materials. The electrical conductivity of pristine epoxy is reported to be on the order of $\sim 10^{-16}\ \mathrm{Sm}^{-1}$ [87], which is below the detection limit of the equipment used in in this study. The addition of only 5 wt% to 7 wt% of graphene, increases the electrical conductivity by ~5 to ~10 orders of magnitudes, respectively. The abrupt change in the electrical conductivity of the epoxy as a result of adding electrically conductive fillers is conventionally described by the power scaling law as $\sigma \sim (\phi - \phi_E)^t$ where $\phi$ is the filler *volume* loading fraction, $\phi_E$ is the filler *volume* fraction at the electrical percolation threshold, and $t$ is the "universal exponent"[88]. We fitted the experimental data in Figure 4 (a) with the power law (dashed lines). The filler loading at the electrical percolation threshold was extracted to be $\phi_E \sim 2.6$ vol%, which corresponds to $f_E \sim 7$ wt%. The extracted universal exponent is $t = 4.2$. The electrical percolation in the composites with various carbon fillers has been investigated extensively, and the values derived in this study agree well with the prior reports[45,88–90]. In the epoxy composites with a single type of filler, the loading fraction at which the thermal percolation is achieved is usually larger than that of the filler loading required to obtain the electrical percolation *i.e.* thermal percolation happens after electrical percolation. In a few studies, the thermal conductivity of the composites did not exhibit changes expected at the percolation[91]. This is because matrix materials such as epoxy, while being completely electrically insulating, still conduct heat. The intrinsic electrical conductivity of the fillers is usually ~15 orders of magnitude larger than that of the polymer matrix while the thermal conductivity is ~2 – 5 orders of magnitude larger than that of the matrix. Because of the high contrast in the electrical conductivity of the fillers and the base polymer matrix, the formation of even a few electrically conductive percolation networks of attached fillers leads to a strong enhancement of the composite electrical conductivity. The formation of a few electrically and thermally conductive pathways does not necessarily result in a major change in the thermal conductivity.

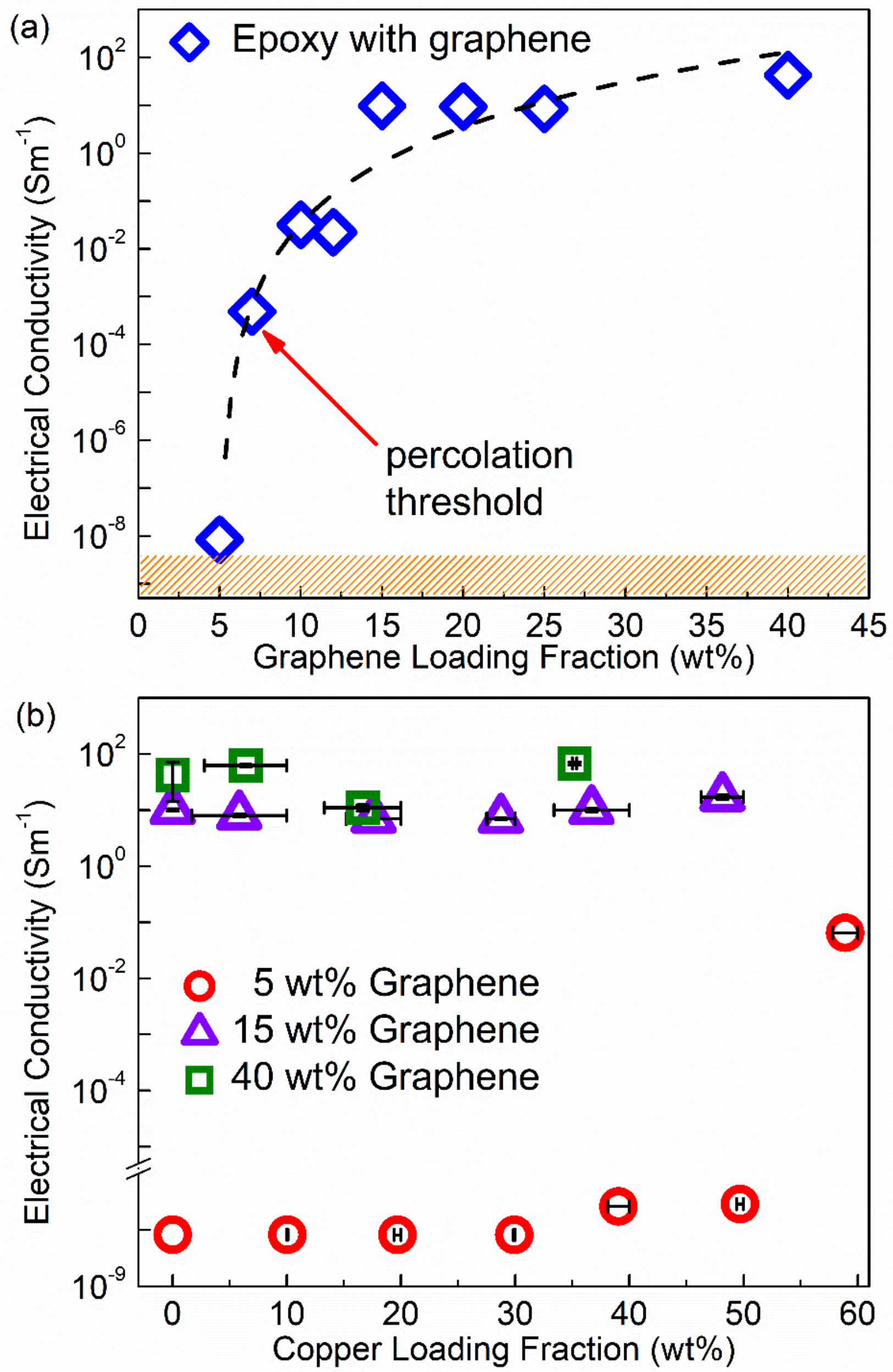

**Figure 4:** (a) Electrical conductivity of the epoxy composites as a function of the graphene filler concentration. The abrupt change in the electrical conductivity, known as the electrical percolation, is achieved at ~7 wt% of the graphene loading fraction. The orange dashed area indicates the instrumental limitation for the electrical conductivity measurements. (b) Electrical conductivity of the epoxy composites with different copper loading fractions at the constant graphene concentrations. The composites with 5 wt% of graphene are highly resistive to the electrical current and addition of copper does not affect the electrical properties. At 60 wt% of copper loading, the electric conductivity increases by six orders of magnitude, revealing the creation of electrically conductive pathways. In the composites with 15 wt% and 40 wt% of graphene, the addition of copper does not produce any effect on the electrical characteristics.

Figure 4 (b) presents the electrical conductivity of the composites at low, medium, and high graphene loadings as a function of the Cu-NP concentration. At the low graphene loading of $f_g = 5$ wt%, addition of the Cu-NP fillers does not affect the electrical characteristics over a wide range of Cu-NP concentrations. However, at $f_{cu}$~60 wt%, the electrical conductivity abruptly grows by ~7 orders of magnitude, confirming the transition to the electrical percolation transport regime. Surprisingly, in this specific case, the electrical percolation threshold is achieved after the thermal percolation threshold, which happens at $f_{cu}$~40 wt% (see Figure 3 (b)). Achieving the electrical percolation after thermal percolation provides the opportunity to prepare electrically insulating TIMs with the high thermal conductivity. For the medium and high graphene loading fractions, adding Cu-NP fillers does not introduce much difference since the composite is already in the electrical percolation regime with the FLG fillers alone. As seen in Figure 4 (a), the composite with the large graphene fillers undergo transition to the electrical percolation regime at $f_g$~7 wt%.

It is interesting to estimate the contribution of electrons and phonons to thermal transport in the low, medium, and high graphene loading composites, at various Cu-NP concentrations. In Cu-NP, like other metals, $\lambda_e$ is the dominant contribution to heat conduction owning to the large concentration of the free electrons. The average diameter of Cu-NPs utilized in this study is ~100 nm, which is larger than the electron MFP, which is ~40 nm. One can assume that the thermal conductivity of the metal particles does not degrade substantially due to the electron – boundary scattering, and remains close to $\sim 400\ \mathrm{Wm^{-1}K^{-1}}$ at RT. In graphene, on the other hand, heat conduction is dominated by phonons[16]. In a complex system of epoxy with metallic and graphene fillers, both types of heat carriers can potentially contribute to thermal transport. We extract $\lambda_e$ from the Wiedemann – Franz law, $\lambda_e/(\sigma T) = (\pi^2/3)(k_B/e)^2$, where $T$ is the absolute temperature, $k_B$ is the Boltzmann constant, and $e$ is the charge of an electron, respectively. The highest electrical conductivity of the samples was measured to be $\sigma \sim 100\ \mathrm{Sm^{-1}}$ for the composite with $f_g = 40$ wt% and $f_{Cu}$~8 wt%. From the Wiedemann-Franz law, one finds the electronic thermal conductivity to be $\lambda_e \sim 0.0007\ \mathrm{Wm^{-1}K^{-1}}$. The obtained value is negligible compared to the total thermal conductivity, confirming that the heat is carried primarily by phonons. Strictly speaking, the concept of phonons breaks apart in the disordered material systems. However, one can argue that it is still can be used for acoustic phonons in the elastic medium in the context of heat conduction.

## 3. Conclusions

We reported thermal properties of the epoxy-based hybrid composites with graphene and copper nanoparticle fillers. It was found that the thermal conductivity of composites with a moderate graphene concentration of $f_g = 15$ wt% exhibits an abrupt increase as the loading of copper nanoparticles approaches $f_{Cu} \sim 40$ wt. %, followed by saturation. In contrast, in composites with a high graphene concentration, $f_g = 40$ wt%, the thermal conductivity increases linearly with addition of copper nanoparticles. At all concentrations of the fillers, below and above the electrical percolation threshold, the thermal transport is dominated by phonons. The obtained results shed light on the interaction between graphene fillers and copper nanoparticles in the composites, and demonstrate potential of such hybrid epoxy composites for practical applications in the thermal interface materials and adhesives.

## 4. Experimental Section

**Sample Preparation:** The composites samples were prepared by mixing commercially available few-layer graphene (xGnP graphene nanoplatelets, grade H, XG-Sciences), copper nanoparticles (US Research Nanomaterials, Inc.), and off-the-shelf epoxy set (Allied High Tech Products, Inc.). Based on the final composite graphene and copper nanoparticle concentrations, the base resin and Cu-NP were weighed and mixed inside a vacuum box in argon gas atmosphere in order to prevent oxidation of Cu particles. The resin and Cu-NP were stirred manually inside the vacuum box in order to be certain that the nanoparticles are well-dispersed inside the polymer. Afterwards, the mixture was mixed outside the glove box using the high-shear speed mixer (Flacktek Inc.) at 1000 rpm for several minutes. Afterwards, graphene was weighed and added to the homogeneous resin-Cu-NP mixture in 3 or 4 steps. At each step, the solution was mixed with the speed mixer at ~1000 rpm, then mixed with a home-made needle like mixer in order to prevent agglomeration and vacuumed in order to obtain an air-bubble free mixture. The curing agent (Allied High Tech Products, Inc.) was then added in the prescribed mass ratio of 1:12 with respect to the epoxy resin. The mixture was poured inside a mold with 25.6 mm diameter and 1 mm thickness and pressed gently. The samples were left in the oven for ~2 hours at 70 ºC to cure and solidify.

**Mass Density Measurements:** The mass densities of the composite samples were measured using Archimedes principle with an electronic scale (Mettler Toledo). The sample were weighed in two different mediums once in air and once when submerged in the water. The density of the sample at RT can be defined using $\rho_c = (w_a/(w_a - w_w)) \times (\rho_w - \rho_a) + \rho_a$ where $w_a$, $w_w$ are the sample's weight in air and in water and $\rho_w$ and $\rho_a$ are the density of the water and air at RT.

**Thermal Diffusivity, Specific Heat Capacity, and Thermal Conductivity Measurements:** Thermal diffusivity of the composites were measured using the transient "laser flash" (LFA) technique (NETZSCH LFA 467 HyperFlash) compliant with the international standard ASTM E-1461, DIM EN 821 and DIN 30905. In this method, the bottom surface of the sample is heated via focusing a short light pulse irradiated from a Xenon flash lamp. A high speed detector acquires the temperature rise of the upper side of the sample. The signal received by the detector is amplified and plotted as a function of time. Thermal diffusivity of the sample is calculated based on the thickness and the time which is required for sample's temperature to reach its 50% ultimate value. Specific heat of the sample is calculated based on comparison of the temperature rise of the sample with that of the known reference sample. The thermal diffusivity and heat capacity are used to determine the thermal conductivity via the equation $\lambda = \rho_c c_p \alpha$ where $\rho_c, c_p$, and $\alpha$ are the mass density, specific heat, and thermal diffusivity of the samples, respectively. More details on the measurement procedures can be found in our prior reports on other material systems[9,11,80,81].

**Electrical Resistivity and Conductivity Measurements:** In order to measure the cross-plane electrical resistivity ($\zeta$), two circular large area Ti/Au contacts (15/150 nm thick) were created on both top and bottom sides of the sample. Resistance ($R$) was measured following the standard two-probe measurements using a digital multimeter (Fluke Corp.) as well as a semiconductor device analyzer (Agilent technology B1500A). The contact resistance was negligible, and the total resistance was dominated by the sample. This is partially due to the large surface area and ohmic contact between Ti/Au and the surface of the composite. Resistivity was calculated via the equation $\zeta = RA/t$ where $R$, $A$, and $t$ are the resistance, contact area, and thickness of the composites, respectively. Finally, conductivity was calculated via the equation $\sigma = 1/\zeta$ and plotted as a function of filler loadings for different composites.

***Acknowledgements***

This work was supported, in part, by the National Science Foundation (NSF) through the Emerging Frontiers of Research Initiative (EFRI) 2-DARE award 1433395: Novel Switching Phenomena in Atomic $MX_2$ Heterostructures for Multifunctional Applications, and by the UC – National Laboratory Collaborative Research and Training Program. D.C. and L.M. acknowledge the support of the National Science Foundation (NSF) through the CAREER award 1351386.

**Contributions**

A.A.B. and F.K. conceived the idea of the study. A.A.B. coordinated the project and contributed to the experimental and theoretical data analysis; Z.B. prepared the composites and performed thermal and electrical conductivity measurements, sample characterization, and contributed to the data analysis; F.K. conducted data analysis and assisted with the thermal measurements; A.M. and C.T.H. contributed to the sample characterization; A.G. conducted the electrical conductivity measurements; D.C. contributed to the sample preparation; L.M. assisted with data analysis; F.K. and A.A.B. led the manuscript preparation. All authors contributed to writing and editing of the manuscript.

## References

[1] S. Ghosh, I. Calizo, D. Teweldebrhan, E. P. Pokatilov, D. L. Nika, A. A. Balandin, W. Bao, F. Miao, C. N. Lau, *Appl. Phys. Lett.* **2008**, *92*, 151911.

[2] J. Renteria, S. Legedza, R. Salgado, M. P. P. Balandin, S. Ramirez, M. Saadah, F. Kargar, A. A. Balandin, *Mater. Des.* **2015**, *88*, 214.

[3] A. Bar-Cohen, K. Matin, S. Narumanchi, *J. Electron. Packag.* **2015**, *137*, 40803.

[4] A. Christensen, S. Graham, *Appl. Therm. Eng.* **2009**, *29*, 364.

[5] P. Goli, S. Legedza, A. Dhar, R. Salgado, J. Renteria, A. A. Balandin, *J. Power Sources* **2014**, *248*, 37.

[6] E. Lee, R. A. Salgado, B. Lee, A. V. Sumant, T. Rajh, C. Johnson, A. A. Balandin, E. V. Shevchenko, *Carbon* **2018**, *129*, 702.

[7] M. Saadah, E. Hernandez, A. A. Balandin, *Appl. Sci.* **2017**, *7*, 589.

[8] X. Xie, D. Li, T.-H. Tsai, J. Liu, P. V. Braun, D. G. Cahill, *Macromolecules* **2016**, *49*, 972.

[9] R. Prasher, *Proc. IEEE* **2006**, *94*, 1571.

[10] K. M. F. Shahil, A. A. Balandin, *Nano Lett.* **2012**, *12*, 861.

[11] K. S. Novoselov, A. K. Geim, S. V. Morozov, D. Jiang, M. I. Katsnelson, I. V. Grigorieva, S. V. Dubonos, A. A. Firsov, *Nature* **2005**, *438*, 197.

[12] Y. Zhang, Y.-W. Tan, H. L. Stormer, P. Kim, *Nature* **2005**, *438*, 201.

[13] R. R. Nair, P. Blake, A. N. Grigorenko, K. S. Novoselov, T. J. Booth, T. Stauber, N. M. R. Peres, A. K. Geim, *Science* **2008**, *320*, 1308.

[14] A. A. Balandin, S. Ghosh, W. Bao, I. Calizo, D. Teweldebrhan, F. Miao, C. N. Lau, *Nano Lett.* **2008**, *8*, 902.

[15] S. Ghosh, I. Calizo, D. Teweldebrhan, E. P. Pokatilov, D. L. Nika, A. A. Balandin, W.

Bao, F. Miao, C. N. Lau, *Appl. Phys. Lett.* **2008**, *92*, 151911.

[16] A. A. Balandin, *Nat. Mater.* **2011**, *10*, 569.

[17] S. Ghosh, W. Bao, D. L. Nika, S. Subrina, E. P. Pokatilov, C. N. Lau, A. A. Balandin, *Nat. Mater.* **2010**, *9*, 555.

[18] J. H. Seol, I. Jo, A. L. Moore, L. Lindsay, Z. H. Aitken, M. T. Pettes, X. Li, Z. Yao, R. Huang, D. Broido, N. Mingo, R. S. Ruoff, L. Shi, *Science* **2010**, *328*, 213.

[19] W. Cai, A. L. Moore, Y. Zhu, X. Li, S. Chen, L. Shi, R. S. Ruoff, *Nano Lett.* **2010**, *10*, 1645.

[20] L. A. Jauregui, Y. Yue, A. N. Sidorov, J. Hu, Q. Yu, G. Lopez, R. Jalilian, D. K. Benjamin, D. A. Delk, W. Wu, Z. Liu, X. Wang, Z. Jiang, X. Ruan, J. Bao, S. S. Pei, Y. P. Chen, In *ECS Transactions*; The Electrochemical Society, 2010; Vol. 28, pp. 73–83.

[21] P. G. Klemens, D. F. Pedraza, *Carbon* **1994**, *32*, 735.

[22] G. Fugallo, A. Cepellotti, L. Paulatto, M. Lazzeri, N. Marzari, F. Mauri, *Nano Lett.* **2014**, *14*, 6109.

[23] D. L. Nika, S. Ghosh, E. P. Pokatilov, A. A. Balandin, *Appl. Phys. Lett.* **2009**, *94*, 203103.

[24] F. Kargar, Z. Barani, R. Salgado, B. Debnath, J. S. Lewis, E. Aytan, R. K. Lake, A. A. Balandin, *ACS Appl. Mater. Interfaces* **2018**, *10*, 37555.

[25] K. M. F. Shahil, A. a. Balandin, *Solid State Commun.* **2012**, *152*, 1331.

[26] Z. Yan, G. Liu, J. M. Khan, A. A. Balandin, *Nat. Commun.* **2012**, *3*, 827.

[27] P. Goli, H. Ning, X. Li, C. Y. Lu, K. S. Novoselov, A. A. Balandin, *Nano Lett.* **2014**, *14*, 1497.

[28] F. Kargar, R. Salgado, S. Legedza, J. Renteria, A. A. Balandin, In *Proc. SPIE*; 2014; Vol. 9168, pp. 91680S-91680S–5.

[29] F. Kargar, Z. Barani, M. Balinskiy, A. S. Magana, J. S. Lewis, A. A. Balandin, *Adv.*

*Electron. Mater.* **2018**, 1800558.

[30] Y.-X. Fu, Z.-X. He, D.-C. Mo, S.-S. Lu, *Int. J. Therm. Sci.* **2014**, *86*, 276.

[31] M. Shtein, R. Nadiv, M. Buzaglo, O. Regev, *ACS Appl. Mater. Interfaces* **2015**, *7*, 23725.

[32] S. Choi, J. Kim, *Compos. Part B Eng.* **2013**, *51*, 140.

[33] J. Gu, Y. Guo, X. Yang, C. Liang, W. Geng, L. Tang, N. Li, Q. Zhang, *Compos. Part A Appl. Sci. Manuf.* **2017**, *95*, 267.

[34] T.-L. Li, S. L.-C. Hsu, *J. Phys. Chem. B* **2010**, *114*, 6825.

[35] A. Yu, P. Ramesh, X. Sun, E. Bekyarova, M. E. Itkis, R. C. Haddon, *Adv. Mater.* **2008**, *20*, 4740.

[36] R. Kumar, S. K. Nayak, S. Sahoo, B. P. Panda, S. Mohanty, S. K. Nayak, *J. Mater. Sci. Mater. Electron.* **2018**, 1.

[37] M. Shtein, R. Nadiv, M. Buzaglo, K. Kahil, O. Regev, *Chem. Mater.* **2015**, *27*, 2100.

[38] T. M. L. Dang, C.-Y. Kim, Y. Zhang, J.-F. Yang, T. Masaki, D.-H. Yoon, *Compos. Part B Eng.* **2017**, *114*, 237.

[39] Y.-K. Kim, J.-Y. Chung, J.-G. Lee, Y.-K. Baek, P.-W. Shin, *Compos. Part A Appl. Sci. Manuf.* **2017**, *98*, 184.

[40] Y. Hernandez, V. Nicolosi, M. Lotya, F. M. Blighe, Z. Sun, S. De, I. T. McGovern, B. Holland, M. Byrne, Y. K. Gun'Ko, J. J. Boland, P. Niraj, G. Duesberg, S. Krishnamurthy, R. Goodhue, J. Hutchison, V. Scardaci, A. C. Ferrari, J. N. Coleman, *Nat. Nanotechnol.* **2008**, *3*, 563.

[41] M. Lotya, Y. Hernandez, P. J. King, R. J. Smith, V. Nicolosi, L. S. Karlsson, F. M. Blighe, S. De, Z. Wang, I. T. McGovern, G. S. Duesberg, J. N. Coleman, *J. Am. Chem. Soc.* **2009**, *131*, 3611.

[42] M. J. Allen, V. C. Tung, R. B. Kaner, *Chem. Rev.* **2010**, *110*, 132.

[43] C. Nethravathi, M. Rajamathi, *Carbon* **2008**, *46*, 1994.

[44] S. Stankovich, D. A. Dikin, R. D. Piner, K. A. Kohlhaas, A. Kleinhammes, Y. Jia, Y. Wu, S. B. T. Nguyen, R. S. Ruoff, *Carbon* **2007**, *45*, 1558.

[45] J. R. Potts, D. R. Dreyer, C. W. Bielawski, R. S. Ruoff, *Polymer* **2011**, *52*, 5.

[46] D. Gall, *J. Appl. Phys.* **2016**, *119*, 085101.

[47] M. S. Liu, M. C. C. Lin, I. Te Huang, C. C. Wang, *Chem. Eng. Technol.* **2006**, *29*, 72.

[48] W. J. Parker, R. J. Jenkins, C. P. Butler, G. L. Abbott, *J. Appl. Phys.* **1961**, *32*, 1679.

[49] P. S. Gaal, M.-A. Thermitus, D. E. Stroe, *J. Therm. Anal. Calorim.* **2004**, *78*, 185.

[50] J. C. M. Garnett, Colours in Metal Glasses, in Metallic Films, and in Metallic Solutions. II. *Philos. Trans. R. Soc. London. Ser. A, Contain. Pap. a Math. or Phys. Character* **1906**, *205*, 237–288.

[51] K. Pietrak, T. S. Winiewski, *J. Power Technol.* **2015**, *95*, 14.

[52] T. L. Bergman, F. P. Incropera, A. S. Lavine, D. P. DeWitt, *Introduction to heat transfer*; John Wiley & Sons, 2011.

[53] D. L. Nika, A. A. Balandin, *Reports Prog. Phys.* **2017**, *80*, 036502.

[54] C.-W. Nan, G. Liu, Y. Lin, M. Li, *Appl. Phys. Lett.* **2004**, *85*, 3549.

[55] P. Bujard, *Intersoc. Conf. Therm. Phenom. Fabr. Oper. Electron. Components I-THERM '88* **1988**, 41.

[56] Y. Zhou, X. Zhuang, F. Wu, F. Liu, Y. Zhou, X. Zhuang, F. Wu, F. Liu, *Crystals* **2018**, *8*, 398.

[57] W. Yu, H. Xie, L. Chen, M. Wang, W. Wang, *Polym. Compos.* **2017**, *38*, 2221.

[58] L. Yu, J. S. Park, Y.-S. Lim, C. S. Lee, K. Shin, H. J. Moon, C.-M. Yang, Y. S. Lee, J. H. Han, *Nanotechnology* **2013**, *24*, 155604.

[59] X. Cui, P. Ding, N. Zhuang, L. Shi, N. Song, S. Tang, *ACS Appl. Mater. Interfaces* **2015**, *7*, 19068.

[60] J. Yu, H. K. Choi, H. S. Kim, S. Y. Kim, *Compos. Part A Appl. Sci. Manuf.* **2016**, *88*, 79.

[61] Y. Feng, X. Li, X. Zhao, Y. Ye, X. Zhou, H. Liu, C. Liu, X. Xie, *ACS Appl. Mater. Interfaces* **2018**, *10*, 21628.

[62] R. Sun, H. Yao, H.-B. Zhang, Y. Li, Y.-W. Mai, Z.-Z. Yu, *Compos. Sci. Technol.* **2016**, *137*, 16.

[63] L. Chen, P. Zhao, H. Xie, W. Yu, *Compos. Sci. Technol.* **2016**, *125*, 17.

[64] L. Shao, L. Shi, X. Li, N. Song, P. Ding, *Compos. Sci. Technol.* **2016**, *135*, 83.

[65] B. Zhao, S. Wang, C. Zhao, R. Li, S. M. Hamidinejad, Y. Kazemi, C. B. Park, *Carbon* **2018**, *127*, 469.

[66] C. Liu, C. Chen, H. Wang, M. Chen, D. Zhou, Z. Xu, W. Yu, *Polym. Compos.* **2018**.

[67] F. Du, W. Yang, F. Zhang, C.-Y. Tang, S. Liu, L. Yin, W.-C. Law, *ACS Appl. Mater. Interfaces* **2015**, *7*, 14397.

[68] S.-Y. Yang, W.-N. Lin, Y.-L. Huang, H.-W. Tien, J.-Y. Wang, C.-C. M. Ma, S.-M. Li, Y.-S. Wang, *Carbon* **2011**, *49*, 793.

[69] H. Im, J. Kim, *Carbon* **2012**, *50*, 5429.

[70] W. Yuan, Q. Xiao, L. Li, T. Xu, *Appl. Therm. Eng.* **2016**, *106*, 1067.

[71] C. Zeng, S. Lu, L. Song, X. Xiao, J. Gao, L. Pan, Z. He, J. Yu, *RSC Adv.* **2015**, *5*, 35773.

[72] K. Kim, H. Ju, J. Kim, *Compos. Sci. Technol.* **2016**, *123*, 99.

[73] Y. Yao, J. Sun, X. Zeng, R. Sun, J.-B. Xu, C.-P. Wong, *Small* **2018**, *14*, 1704044.

[74] H. Song, B. Kim, Y. Kim, Y.-S. Bae, J. Kim, Y. Yoo, H. Song, B. G. Kim, Y. S. Kim, Y.-S. Bae, J. Kim, Y. Yoo, *Polymers* **2019**, *11*, 484.

[75] K. Kim, M. Kim, J. Kim, *Compos. Sci. Technol.* **2014**, *103*, 72.

[76] C.-C. Teng, C.-C. M. Ma, K.-C. Chiou, T.-M. Lee, *Compos. Part B Eng.* **2012**, *43*, 265.

[77] S. Y. Pak, H. M. Kim, S. Y. Kim, J. R. Youn, *Carbon* **2012**, *50*, 4830.

[78] F. Wang, X. Zeng, Y. Yao, R. Sun, J. Xu, C.-P. Wong, *Sci. Rep.* **2016**, *6*, 19394.

[79] T. Zhou, X. Wang, X. Liu, D. Xiong, *Carbon* **2010**, *48*, 1171.

[80] A. Ma, W. Chen, Y. Hou, *Polym. Plast. Technol. Eng.* **2012**, *51*, 1578.

[81] P. Zhang, Q. Li, Y. Xuan, *Compos. Part A Appl. Sci. Manuf.* **2014**, *57*, 1.

[82] C. Chen, H. Wang, Y. Xue, Z. Xue, H. Liu, X. Xie, Y.-W. Mai, *Compos. Sci. Technol.* **2016**, *128*, 207.

[83] T. Huang, X. Zeng, Y. Yao, R. Sun, F. Meng, J. Xu, C. Wong, *RSC Adv.* **2016**, *6*, 35847.

[84] F. Ren, D. Song, Z. Li, L. Jia, Y. Zhao, D. Yan, P. Ren, *J. Mater. Chem. C* **2018**, *6*, 1476.

[85] S. D. Rad, A. Islam, A. Alnasser, *J. Compos. Mater.* **2018**, 002199831881292.

[86] Y. Zhou, S. Yu, H. Niu, F. Liu, Y. Zhou, S. Yu, H. Niu, F. Liu, *Polymers* **2018**, *10*, 1412.

[87] S. Singha, M. Thomas, *IEEE Trans. Dielectr. Electr. Insul.* **2008**, *15*, 12.

[88] S. Stankovich, D. A. Dikin, G. H. B. Dommett, K. M. Kohlhaas, E. J. Zimney, E. A. Stach, R. D. Piner, S. B. T. Nguyen, R. S. Ruoff, *Nature* **2006**, *442*, 282.

[89] M. J. Biercuk, M. C. Llaguno, M. Radosavljevic, J. K. Hyun, A. T. Johnson, J. E. Fischer, *Appl. Phys. Lett.* **2002**, *80*, 2767.

[90] K. Awasthi, S. Awasthi, A. Srivastava, R. Kamalakaran, S. Talapatra, P. M. Ajayan, O. N. Srivastava, *Nanotechnology* **2006**, *17*, 5417.

[91] N. Shenogina, S. Shenogin, L. Xue, P. Keblinski, *Appl. Phys. Lett.* **2005**, *87*, 1.